\documentclass[]{article}

\usepackage{amsmath}
\usepackage{hyperref}
\usepackage{cleveref}
\usepackage{graphicx}

\usepackage{biblatex}

\title{Suppressing Spontaneous Droplet Shrinkage in Cahn-Hilliard-Stokes Microflows}

\author{A. M. Gurin\thanks{Lavrentyev Institute of Hydrodynamics, SB RAS, Russia, Novosibirsk.}}

\begin{document}

\maketitle

\begin{abstract}
This paper addresses non-physical artifact, specifically spontaneous droplet shrinkage, in Cahn-Hilliard-based simulations of immiscible two-phase microflows. That artifact arise from the energy functional’s preference for reduced interfacial area, leading to shifts in equilibrium phase fractions away from the physical bounds of 0 and 1. A curvature-dependent correction is proposed that shifts the double-well potential to counteract this drift. This method is validated on the following test cases: an isolated droplet in a cubic cavity, drainage in a synthetic porous medium, and flow through a T-shaped junction. Results demonstrate significant suppression of phase fraction overshoot/undershoot and improved droplet volume conservation.
\end{abstract}

\section{Introduction}

Efficient oil extraction relies heavily on reservoir-scale simulation software. These modeling tools enable users to evaluate economic viability and optimize the extraction process. However, the accuracy of such software depends on reliable estimates of the permeability of porous media. Traditionally, these values are obtained through laboratory experiments on rock core samples extracted from reservoir depths. This approach is costly, and the samples are often destroyed during testing, preventing their reuse. As a result, Digital Rock Physics (DRP) has emerged as an alternative approach, allowing engineers to simulate multiphase flow within accurate representations of rock pore structures. DRP combines microtomographic imaging of subsurface rock with numerical simulations of multiphase flow. Among the available numerical methods, the phase-field approach, and in particular the Cahn--Hilliard--Stokes system, is widely used for modeling multiphase flow in porous media. Unlike other approaches, such as pore-network models or the Volume of Fluid (VOF) method, it is based on the minimization of a free-energy functional. Although this method offers significant advantages, it also suffers from a well-known limitation related to the formulation of the energy functional: in particular, the standard Cahn--Hilliard equation may fail to preserve the apparent droplet volume accurately, even though the total phase mass is conserved.

Problems associated with droplet shrinkage have been known for a long time. The paper \cite{YUE20071} shows that this problem is related to the system energy functional and is not merely a numerical effect. The authors show that the energy functional makes droplet compression energetically favorable. Compression of a droplet reduces its surface area and, consequently, its energy. The droplet continues to shrink until this decrease is compensated by an increase in energy caused by a change in the phase fraction inside the droplet. From these conditions, the authors derived the value of $\delta \phi$ by which the phase fraction exceeds its physical bounds. The paper \cite{YUE20071} proposes reducing the interfacial width and choosing the mobility so that non-physical effects are significantly reduced. An important part of the paper is the derivation of a theoretical basis that links droplet disappearance with changes in the equilibrium values of the phase fraction outside and inside the droplet. The study \cite{Drop_shrink_on_surface} similarly considers droplet shrinkage on a surface with a fixed contact angle, rather than in free space.

There are more effective methods for mitigating droplet shrinkage than those proposed in earlier studies. In \cite{Lagrange_Multip_CH}, a Lagrange multiplier method is used to constrain the phase fraction and prevent it from exceeding prescribed bounds. The examples presented in that study show an ideal distribution of the phase fraction between 0 to 1. However, the use of Lagrange multipliers significantly complicates the governing equations and makes them more difficult to solve. In \cite{Ostwald_Log_potential_CH}, a logarithmic form of the double-well potential is used, which has a steeper slope near the nonphysical region. This potential greatly reduces nonphysical effects, but the logarithmic singularity also introduces difficulties in solving the equations. These two methods prevent the phase fraction from leaving the admissable interval from 0 to 1, however, as will be shown later in this paper, the phase fraction can also deviate from its desired equilibrium values while still remaining within these bounds.

There are also methods for iteratively compensating for phase-fraction overshoot. These methods require a split-method approach to solving the equations. In \cite{Drop_shrink_compensation_1}, after solving the Cahn--Hilliard equations, the phase fractions that have exceeded the physical bounds are redistributed. Since there is only a single-phase droplet on a substrate, the excess phase fraction is uniformly distributed across the entire interfacial boundary. In the case of multiple droplets, the method proposed in \cite{Drop_shrink_compensation_2} can be applied. This method uses the discontinuous Galerkin method to solve the Cahn--Hilliard--Stokes equations, and phase-fraction overshoot is prevented by limiting the phase fluxes.

In this paper, the Cahn--Hilliard--Stokes equations are formulated, and an implicit finite-difference numerical scheme on a staggered grid is presented. Within this model, the problem of droplet shrinkage is discussed and demonstrated using synthetic test cases: a T-shaped junction and a phase droplet in a box cavity. A correction for the two-phase model that significantly suppresses non-physical effects is proposed. The correction is based on shifting the double-well potential by an amount that depends on the curvature of the interfacial boundary, obtained by analogy with \cite{YUE20071}. The modified equations are tested on synthetic problems and on drainage through a porous medium.

\section{Mathematical model}

We consider the dynamics of a two-phase fluid system evolving within a complex, fully resolved micropore network. The hydrodynamics are governed by the non-stationary incompressible creeping flow approximation, where the fluid motion is propelled by a macroscopic pressure difference and opposed by viscous dissipation and capillary force. The liquid mixture consists of two distinct, incompressible, and immiscible components.
To capture the topological changes and interfacial dynamics without explicit interface tracking, the phase distribution is modeled using a continuous order parameter(phase volume fraction). Consequently, the computational domain is divided into bulk regions of Fluid 1 and Fluid 2, separated by a diffuse interfacial region of finite thickness. In this formulation, surface tension effects manifest as capillary forces driven by gradients in phase fraction. Physically, this results in a pressure jump that acts to minimize interfacial area, pushing the interface from the convex side toward the concave side. Interaction with the solid matrix is incorporated through wetting boundary conditions, where the preferential affinity of the fluids to the pore walls is prescribed via a contact angle $\theta$.


We consider a phase-field model based on the Cahn--Hilliard equation with a double-well potential \cite{Cahn_Hilliard_two_phase_0_1}. The phase distribution over the computational domain is represented by the phase field variable \(\phi\), which supposed to take values \(\phi = 0\) and \(\phi = 1\) in the two pure phases and varies smoothly between these values across the interfacial region. The interface has a finite thickness, controlled by the parameter \(\varepsilon\).

The model is derived from the minimization of the Ginzburg--Landau free energy functional \(F\):
\begin{equation} \label{energy functional two phase}
F(\phi, \nabla \phi) := \int\limits_\Omega \left ( 12 \frac{\sigma}{\varepsilon} \Psi(\phi) + \frac{3}{4} \sigma \varepsilon |\nabla \phi|^2 \right ) d\Omega,
\end{equation}
where
\begin{equation} \label{two_well_potential}
\Psi(\phi) = \phi^2 (1-\phi)^2.
\end{equation}
Here, \(\Psi(\phi)\) represents the bulk (or volumetric) contribution to the energy, while \(|\nabla \phi|^2\) corresponds to the interfacial (gradient) contribution. The constant \(\sigma\) denotes the surface tension coefficient. The bulk energy density \(\Psi(\phi)\) attains its minima at \(\phi = 0\) and \(\phi = 1\), corresponding to the two coexisting phases.

The evolution of the phase field is governed by the Cahn--Hilliard equation:

\begin{equation*}
\frac{\partial \phi}{\partial t} - \nabla \cdot ( M \nabla \mu ) + \nabla \cdot \left ( \vec{u} \phi \right ) = 0,
\end{equation*}
where the chemical potential $\mu$ is defined as:
\begin{equation} \label{chem_potential_eq}
\mu = 12 \frac{\sigma}{\varepsilon} \Psi'(\phi) - \frac{3}{2} \sigma \varepsilon \Delta \phi,
\end{equation}
with the derivative of the bulk energy potential given by:
\begin{equation*}
\Psi'(\phi) = 2 \phi (1 - \phi)(1 - 2 \phi).
\end{equation*}
Here, \(M\) denotes the mobility, \(\mu\) is the chemical potential, and \(\vec{u}\) represents the fluid velocity field. The term \(\nabla \cdot \left ( \vec{u} \phi \right )\) accounts for advection of the phase field by the flow.

The fluid motion is described by the incompressible Stokes equations:
\begin{equation*} 
\nabla \cdot \vec{u} = 0,
\end{equation*}

\begin{equation*}
\frac{\partial \rho \vec{u}}{\partial t} + \nabla p - \nabla \cdot \left (\eta \left ( \nabla \vec{u} + \nabla \vec{u}^T \right ) \right ) - \mu \nabla \phi = 0,
\end{equation*}
where \(\rho\) is the fluid density, \(p\) is the pressure, \(\eta\) is the dynamic viscosity, and the term \(\mu \nabla \phi\) represents the surface tension force arising from interfacial energy gradients.

The density and dynamic viscosity are phase-dependent properties interpolated linearly according to the following expressions:
\begin{equation*}
    \rho(\phi) = \rho_0 \left( 1 - \phi \right) + \rho_1 \phi,
\end{equation*}
\begin{equation*}
    \eta(\phi) = \eta_{0} \left ( 1 - \phi \right ) + \eta_{1} \phi,
\end{equation*}
where subscripts \(0\) and \(1\) denote the properties in the phase corresponding to \(\phi=0\) and \(\phi=1\), respectively.

Boundary conditions are imposed as follows. On solid impermeable walls, the contact angle condition is prescribed via a Neumann-type boundary condition for the phase field:
\begin{equation*} 
    \nabla \phi \cdot \vec{n} = -\frac{4}{\varepsilon} \text{cos}(\theta_{12}) \phi (1 - \phi),
\end{equation*}
where $\theta_{12}$ is the contact angle between phases 1 and 2, $\vec{n}$ is the outward normal to the impermeable surface.

For the chemical potential \(\mu\), homogeneous Neumann boundary conditions are applied on all boundaries (walls, inlet, and outlet):

$$\nabla \mu \cdot \vec{n} = 0,$$
which ensures mass conservation and avoids spurious phase fluxes through boundaries.

At inflow and outflow boundaries, the phase field is typically advected with the flow. To avoid nonphysical oscillations or excessive diffusion near inlet boundary, a third-order condition is employed:
\begin{equation} \label{third_order_BC}
\nabla \phi \cdot \vec{n} = \frac{\phi_{\text{inlet}} - \phi}{h},
\end{equation}
where \(h\) denotes a mesh cell size and \(\phi_{\text{inlet}}\) is the prescribed inlet phase value. This condition effectively relaxes \(\phi\) toward \(\phi_{\text{inlet}}\) in cells adjacent to the inlet.

For the Stokes system, no-slip boundary conditions (\(\vec{u} = \vec{0}\)) are imposed on impermeable walls. At inlet and outlet boundaries, the velocity is prescribed as uniform profile. The pressure is subject to a natural (homogeneous Neumann) condition, \(\nabla p \cdot \vec{n} = 0\), at all boundaries. This introduces a nullspace in the discrete pressure system (constant pressure mode).

\section{Numerical scheme}

The governing equations are discretized using the finite difference method on a uniform Cartesian grid with cubic cells of size $h$. To correctly couple pressure and velocity, we employ a staggered-grid discretization (MAC grid). Cell-centered quantities are indexed by $(i,j,k)$, while the velocity components are located at the corresponding cell faces: $u_{i+\frac{1}{2},j,k}$, $v_{i,j+\frac{1}{2},k}$, and $w_{i,j,k+\frac{1}{2}}$. For brevity, we describe only the discretization of the $x$-momentum equation; the remaining components are treated analogously.

For a cell-centered scalar field $q$, the standard discrete Laplacian is defined as
\begin{equation*}
\begin{split}
\left(\Delta_h q\right)_{i,j,k}
:=
\frac{q_{i+1,j,k}-2q_{i,j,k}+q_{i-1,j,k}}{h^2}
+
\frac{q_{i,j+1,k}-2q_{i,j,k}+q_{i,j-1,k}}{h^2}
\\
+
\frac{q_{i,j,k+1}-2q_{i,j,k}+q_{i,j,k-1}}{h^2}.
\end{split}
\end{equation*}
The discrete divergence operator at the cell center $(i,j,k)$ is given by
\begin{equation*}
\left (\nabla_h \cdot \vec{u}\right )_{i,j,k} := 
\frac{u_{i+\frac12,j,k}-u_{i-\frac12,j,k}}{h} + 
\frac{v_{i,j+\frac12,k}-v_{i,j-\frac12,k}}{h} + 
\frac{w_{i,j,k+\frac12}-w_{i,j,k-\frac12}}{h}.
\end{equation*}
For the viscous term in the $x$-momentum equation, we write
\begin{equation*}
\mathcal{L}_{\eta,h}^{x}(u,v,w)
=
\delta_x \tau_{xx} +
\delta_y \tau_{xy} +
\delta_z \tau_{xz},
\end{equation*}
where the stress components are discretized at the locations required by the staggered grid. Operators $\delta_x$, $\delta_y$ and $\delta_z$ are a first derivatives along x, y, z axes discretized with finite differences on staggered grid. The normal stress is defined at cell centers:
\begin{equation*}
\left(\tau_{xx}\right)_{i,j,k}
=
2 \eta_{i,j,k} \frac{ u_{i+\frac{1}{2},j,k} - u_{i-\frac{1}{2},j,k} }{h}.
\end{equation*}
The shear stresses are defined at cell edges:
\begin{equation*}
\left(\tau_{xy}\right)_{i+\frac{1}{2},j+\frac{1}{2},k}
=
\eta_{i+\frac{1}{2},j+\frac{1}{2},k} \left[
\frac{ u_{i+\frac{1}{2},j+1,k} - u_{i+\frac{1}{2},j,k} }{h} +
\frac{ v_{i+1,j+\frac{1}{2},k} - v_{i,j+\frac{1}{2},k} }{h} \right],
\end{equation*}
\begin{equation*}
\left(\tau_{xz}\right)_{i+\frac{1}{2},j,k+\frac{1}{2}}
=
\eta_{i+\frac{1}{2},j,k+\frac{1}{2}} \left[
\frac{ u_{i+\frac{1}{2},j,k+1} - u_{i+\frac{1}{2},j,k} }{h} +
\frac{ w_{i+1,j,k+\frac{1}{2}} - w_{i,j,k+\frac{1}{2}} }{h}
\right].
\end{equation*}
The discrete divergence of these stresses at the $u$-velocity location $\left(i+\frac{1}{2},j,k\right)$ is then
\begin{equation*}
\begin{split}
\left(\mathcal{L}_{\eta,h}^{x}\right)_{i+\frac{1}{2},j,k}
=
&
\frac{ \left(\tau_{xx}\right)_{i+1,j,k} - \left(\tau_{xx}\right)_{i,j,k} }{h}
\\
&+
\frac{ \left(\tau_{xy}\right)_{i+\frac{1}{2},j+\frac{1}{2},k} - 
       \left(\tau_{xy}\right)_{i+\frac{1}{2},j-\frac{1}{2},k} }{h}
\\
&+
\frac{ \left(\tau_{xz}\right)_{i+\frac{1}{2},j,k+\frac{1}{2}} -
\left(\tau_{xz}\right)_{i+\frac{1}{2},j,k-\frac{1}{2}} }{h}.
\end{split}
\end{equation*}
The viscosity is stored at cell centers and interpolated to edge locations. For example,
\begin{equation*}
\eta_{i+\frac{1}{2},j+\frac{1}{2},k}
=
\frac{1}{4}
\left( \eta_{i,j,k} + \eta_{i+1,j,k}
+
\eta_{i,j+1,k} + \eta_{i+1,j+1,k}
\right),
\end{equation*}
and
\begin{equation*}
\eta_{i+\frac{1}{2},j,k+\frac{1}{2}}
=
\frac{1}{4}
\left( \eta_{i,j,k} + \eta_{i+1,j,k}
+
\eta_{i,j,k+1} + \eta_{i+1,j,k+1}
\right).
\end{equation*}
The operators $\mathcal{L}_{\eta,h}^{y}$ and $\mathcal{L}_{\eta,h}^{z}$ for the $v$- and $w$-momentum equations are defined analogously.

The surface tension force term and pressure gradient are discretized at the cell faces. For the $x$-component, we use
\begin{equation*}
\left (\mu \delta_x \phi \right )_{i+\frac12,j,k} := 
\frac{\mu_{i+1,j,k} + \mu_{i,j,k}}{2} \cdot \frac{\phi_{i+1,j,k} - \phi_{i,j,k}}{h},
\end{equation*}
and
\begin{equation*}
\left( \delta_x p \right)_{i+\frac12,j,k} = \frac{p_{i+1,j,k} - p_{i,j,k}}{h}.
\end{equation*}
For the advective term in the Cahn--Hilliard equation, a first-order upwind scheme is employed. The numerical flux at the $x$-face is defined as
\begin{equation*}
    \left ( f^x_\phi \right )_{i+\frac12} = \begin{cases} 
        u_{i+\frac12,j,k} \phi_{i,j,k} & \mbox{if } u_{i+\frac12,j,k} \ge 0 \\
        u_{i+\frac12,j,k} \phi_{i+1,j,k} & \mbox{if } u_{i+\frac12,j,k} < 0 
    \end{cases}
\end{equation*}
The fluxes in the $y$- and $z$-directions, $\left( f^y_\phi \right)$ and $\left( f^z_\phi \right)$, are defined analogously using the velocities $v$ and $w$, respectively. The discrete divergence of the flux is then
\begin{equation*}
\left (\nabla_h \cdot \vec{u} \phi \right )_{i,j,k} := 
\frac{\left ( f^x_\phi \right )_{i+\frac12}-\left ( f^x_\phi \right )_{i-\frac12}}{h} + 
\frac{\left ( f^y_\phi \right )_{j+\frac12}-\left ( f^y_\phi \right )_{j-\frac12}}{h} + 
\frac{\left ( f^z_\phi \right )_{k+\frac12}-\left ( f^z_\phi \right )_{k-\frac12}}{h}.
\end{equation*}
Temporal discretization is denoted by:

\begin{equation*}
\delta_t \phi := \frac{\phi^{n+1} - \phi^{n}}{\tau},
\end{equation*}
where $\tau$ is the time step. 

To solve the Cahn--Hilliard equation, we use the energy-stable semi-implicit discretization described in \cite{Stable_scheme_three_phase}. Although originally formulated for a three-phase model, it can be adapted for the two-phase case. This scheme modifies the discretization of the bulk potential derivative $\Psi'$ to ensure $\Psi(\phi^{n+1}) - \Psi(\phi^{n}) = \Psi' \cdot (\phi^{n+1} - \phi^{n})$. The discrete potential derivative is given by:

$$\Psi'(\phi^{n+1},\phi^{n}) = \frac{(\phi_1^{n+1} + \phi_1^n) ( (\phi_2^{n+1})^2 + (\phi_2^{n})^2 ) - (\phi_2^{n+1} + \phi_2^n) ( (\phi_1^{n+1})^2 + (\phi_1^{n})^2 )}{2},$$
where $\phi_1 = \phi$ and $\phi_2 = 1 - \phi$.
The final fully discrete system is summarized as follows:

\begin{equation*}
\delta_t \phi_{i,j,k} - M \left (\Delta_h \mu \right )_{i,j,k} + \left (\nabla_h \cdot \vec{u} \phi\right )_{i,j,k} = 0,
\end{equation*}

\begin{equation*}
\mu_{i,j,k} = 12 \frac{\sigma}{\varepsilon} \Psi'(\phi_{i,j,k}) - \frac{3}{2} \sigma \varepsilon \left (\Delta_h \phi\right )_{i,j,k},
\end{equation*}

\begin{equation*}
\left (\nabla_h \cdot \vec{u}\right )_{i,j,k} = 0,
\end{equation*}

\begin{equation*}
\delta_t \left(\rho u_{i+\frac{1}{2},j,k}\right)
+
\left(\delta_x p\right)_{i+\frac{1}{2},j,k}
-
\left(\mathcal{L}_{\eta,h}^{x}(u,v,w)\right)_{i+\frac{1}{2},j,k}
-
\left(\mu \delta_x \phi\right)_{i+\frac{1}{2},j,k}
=
0.
\end{equation*}
The momentum equations for the $y$- and $z$-components are discretized analogously.

The resulting discrete system of nonlinear equations is linearized using Newton's method. The linear systems arising at each Newton iteration are solved by the FGMRES method with a custom multilevel preconditioner. The numerical scheme described above was implemented within the INMOST framework \cite{INMOST_citation}, which handles computational grids and supports distributed-memory parallelization. The PETSc library \cite{PetSc_citation} is used to solve the systems of equations arising from the discretization and linearization. In particular, PETSc is used to implement a multigrid preconditioner based on \cite{Stokes_Notay_Transformation} that accounts for the specific features of the coupled system.

\section{Flow in a Porous Medium} \label{ch_flow_in_porous_medium}

\begin{figure}[htbp]
\centering
\includegraphics[width=\textwidth]{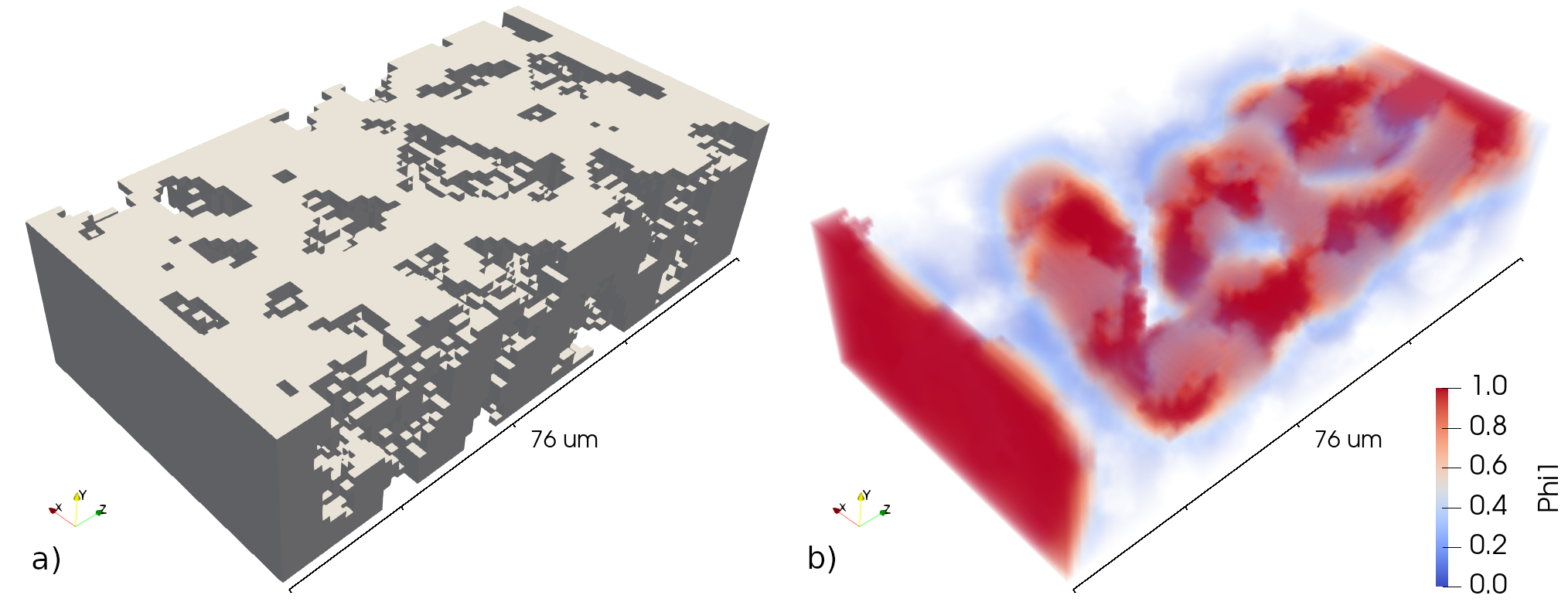}
\caption{(a) Computational grid, (b) initial distribution of the non-wetting phase.}
\label{img:Mesh_initial_distribution}
\end{figure}

Drainage simulations were performed in a porous-media with 66\% porosity. The computational grid was generated using a random porous media generator based on sampling Gaussian noise on several coarser regular grids, followed by tricubic interpolation. The resulting pore-space grid is shown in \cref{img:Mesh_initial_distribution}(a). All cells are cubes with edge length $h = 1\,\mu m$. Three layers of dummy cells were added on each side along the $z$-axis to facilitate the imposition of velocity boundary conditions.

Drainage proceeds along the $z$-axis with a fixed inlet velocity of $1\,mm/s$ at $z = 0\,\mu m$ and a corresponding outlet condition at $z = 76\,\mu m$. For the Cahn--Hilliard equation, we use the mobility $M = 10^{-12}$ and the interface width $\varepsilon = 4\,\mu m$. The two fluids have the following properties: densities $\rho_1 = 800\,kg/m^3$ and $\rho_2 = 1000\,kg/m^3$; dynamic viscosities $\eta_{1} = 0.005\,Pa\cdot s$, $\eta_{2} = 0.001\,Pa\cdot s$; surface tension $\sigma = 0.008\,N/m$; and contact angle $\theta = 0^\circ$, with phase~1 being non-wetting.

The initial phase distribution was prepared by solving the Cahn--Hilliard equation alone, without coupling it to the Stokes equations. A random phase field $\phi \in [0,1]$ was assigned to all cells and evolved until a metastable configuration was reached. \Cref{img:Mesh_initial_distribution}(b) shows the resulting non-wetting-phase distribution, which preferentially occupies larger pores and accumulates near the inlet due to the third-order boundary condition \cref{third_order_BC}. Subsequently, drainage is simulated by injecting 100\% non-wetting phase injected at the inlet.

\begin{figure}[htbp]
\centering
\includegraphics[width=\textwidth]{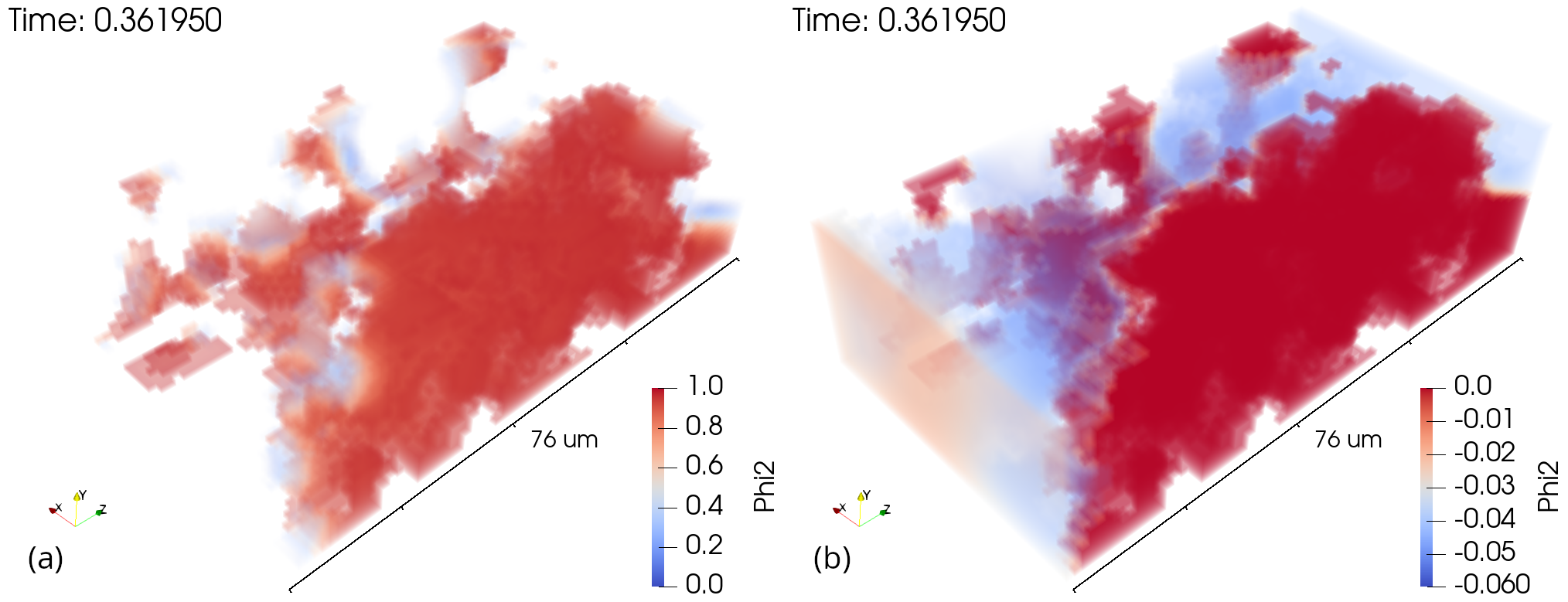}
\caption{Wetting-phase distribution in the porous medium: (a) $\phi$ distribution within the physical bounds, (b) $\phi$ distribution in the range from -0.06 to 0.}
\label{img:porous_med_distribution}
\end{figure}

\begin{figure}[htbp]
\centering
\includegraphics[width=\textwidth]{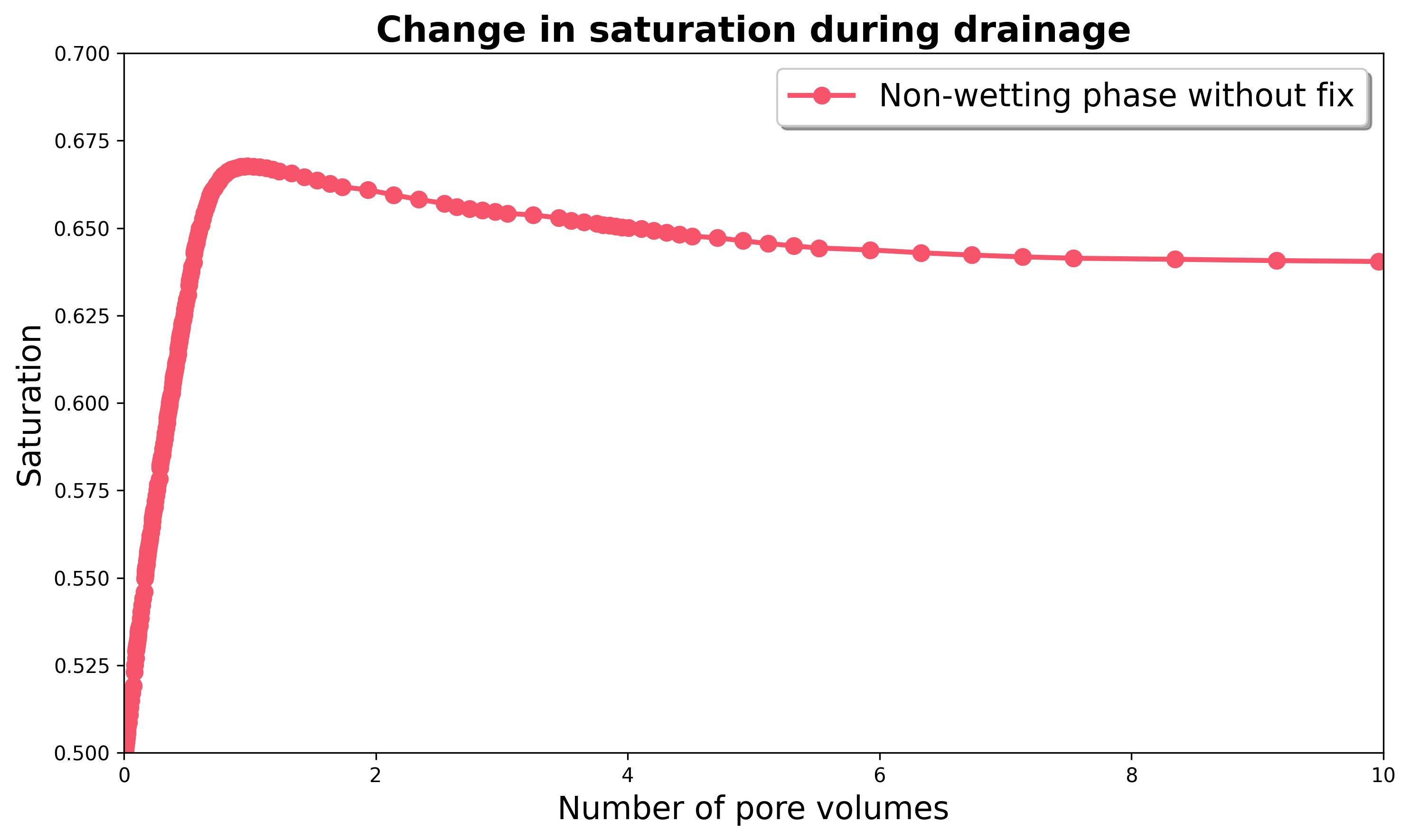}
\caption{Saturation evolution during drainage.}
\label{img:Saturation_plot}
\end{figure}

\Cref{img:porous_med_distribution}(a) shows the wetting-phase $\phi_2 = 1 - \phi$ distribution after the injection of seven pore volumes. A considerable amount of wetting phase remains trapped in the pores. The curve in \cref{img:Saturation_plot} displays the non-wetting-phase saturation, defined as $S = \int_\Omega \phi \, d\Omega$ versus the injected pore volumes. Notably, at approximately 0.7 injected pore volumes, the saturation of the non-wetting-phase begins to decrease. This is a clearly nonphysical effect, since only the non-wetting phase is injected into the computational domain; therefore, its volume should not decrease.

The origin of this nonphysical behavior is clarified in \cref{img:porous_med_distribution}(b), which uses a color scale ranging from $-0.06$ to $0$ for the wetting-phase fraction. Negative wetting-phase values, as low as $-0.04$, leave the domain through the outlet, while the inlet enforces $\phi = 0$. Thus, the wetting phase accumulates in the domain at a rate of approximately 4\% of the pore volume per injected pore volume. In typical drainage simulations, accurately predicting the amount of trapped phase is critical. Such nonphysical phase accumulation can easily change this prediction by a factor of two over three injected pore volumes. Hence, the uncorrected Cahn--Hilliard--Stokes model introduces a systematic error that leads to pronounced nonphysical effects. In the next section, we discuss the intrinsic mechanisms driving the observed behavior.

\section{Droplet shrinkage and dissolution} \label{chap:drop_evaporation}

\begin{figure}[htbp] 
\centering
\includegraphics[width=\textwidth]{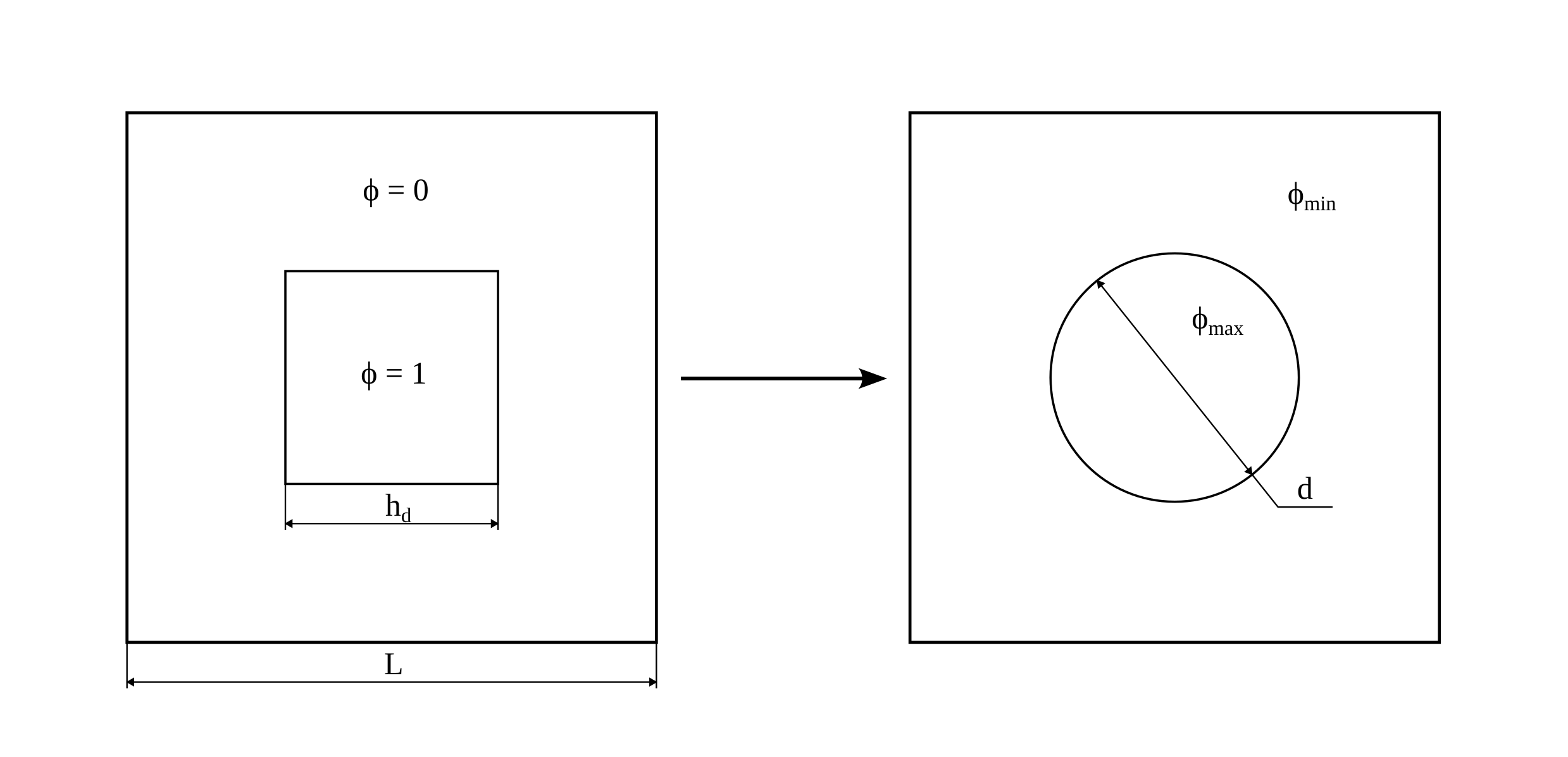}
\caption{Schematic of a droplet in a cubic cavity.}
\label{img:box_scheme}
\end{figure}
To illustrate this nonphysical effects, a series of simulations was performed using the simple geometry shown in \cref{img:box_scheme}. A cubic droplet of one phase with initial edge length $h_d$ is placed at the center of a cubic computational domain filled with the other phase. Initially, the interface is a sharp step between the two phases. As the Cahn--Hilliard equations evolve, the interface diffuses and the droplet assumes a spherical shape. The key parameters are the interface width $\varepsilon$, the domain size $L$, and the initial cube edge length $h_d$. The mobility $M$ and surface tension $\sigma$ influence only the time scale required to reach steady state. They are set to $10^{-11}\,m^5/(J\cdot s)$ and $0.05\,N/m$, respectively. The grid spacing is fixed at $1\,\mu m$ in all simulations, and the interface width is $\varepsilon = 4\,\mu m$.

\begin{figure}[htbp] 
\centering
\includegraphics[width=\textwidth]{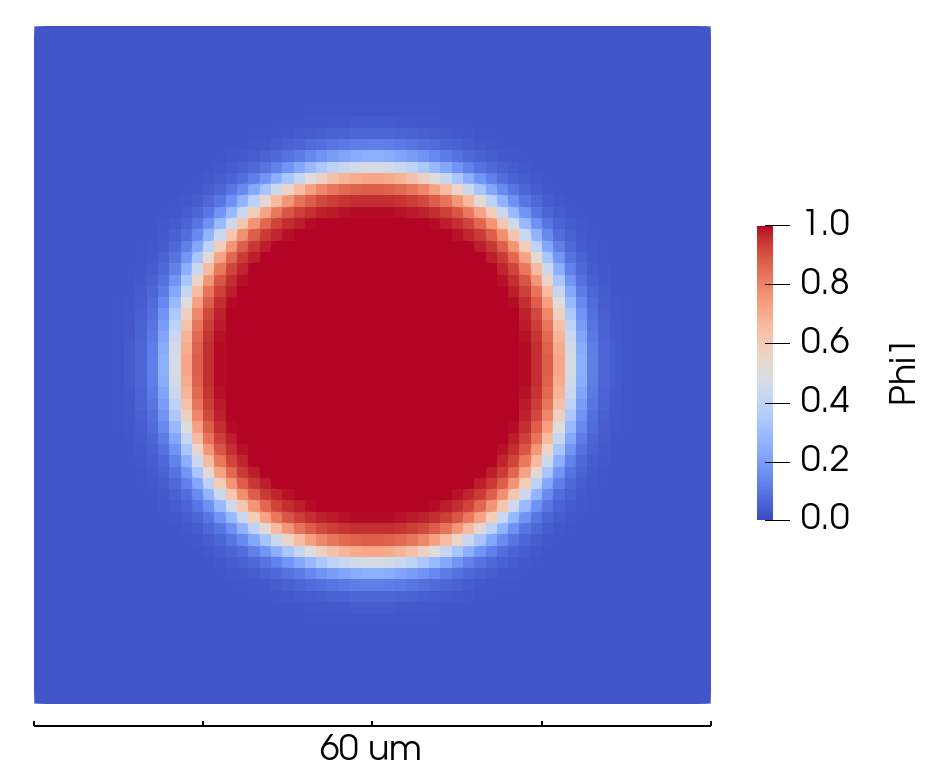}
\caption{Phase-fraction distribution in the computational domain at steady state.}
\label{img:box_60x60}
\end{figure}
An example of a steady-state solution on the center plane of the domain is shown in \cref{img:box_60x60}. A cubic droplet of phase 1 with dimensions $30\times30\times30\,\mu m$ was placed within a $60\times60\times60\,\mu m$ domain. During the evolution, the droplet becomes spherical. The phase fraction reaches constant bulk values inside and outside the droplet, with a smooth transition across approximately eight grid cells at the interface. Although the figure suggests that the steady-state phase fractions are exactly 0 and 1, this is not the case.

\begin{table}[htbp]
\begin{tabular}{|c|c|c|c|c|}
\hline
Case & \(L\), $\mu m$ & \(h_d\), $\mu m$ & \(d\), $\mu m$ & \(\phi_{max}\), \(\phi_{min}\) \\
\hline
1 & 30 & 18 & 19 & 1.02845, 0.035084 \\
2 & 60 & 20 & -- & -- \\
3 & 60 & 25 & 23 & 1.02537, 0.030189 \\
4 & 60 & 30 & 33 & 1.01791, 0.020083 \\
5 & 60 & 35 & 39 & 1.01552, 0.017119 \\
6 & 60 & 40 & 47 & 1.01307, 0.014170 \\
7 & 60 & 45 & -- & -- \\
8 & 80 & 30 & 29 & 1.02049, 0.023419 \\
9 & 80 & 35 & 37 & 1.01638, 0.018171 \\
10 & 80 & 40 & 47 & 1.01338, 0.014558 \\
11 & 80 & 45 & 51 & 1.01205, 0.012985 \\
12 & 80 & 50 & 59 & 1.01053, 0.011236 \\
13 & 80 & 55 & 65 & 1.00973, 0.010320 \\
14 & 80 & 60 & -- & -- \\
\hline
\end{tabular}
\caption{Simulation parameters and results for a droplet in a cubic cavity.}
\label{box_without_fix}
\end{table}
\Cref{box_without_fix} shows the equilibrium phase fractions $\phi_{\max}$, inside the droplet, and $\phi_{\min}$, outside the droplet, as functions of the domain size $L$ and the initial droplet size $h_d$. The equilibrium droplet diameter $d$ is defined as the maximum extent of the region where $\phi > 0.5$. The table reveals that both $\phi_{\max}$ and $\phi_{\min}$ deviate from the ideal values of 1 and 0, with the deviation decreasing as the droplet size increases. The bulk phase fraction inside the droplet is greater than one, implying that the phase fraction of the second phase inside the droplet is negative. In case~2 the droplet fully dissolved and was uniformly smeared over the entire computational domain. In cases~7 and~14, the droplet interacted with the boundaries of the computational domain, and these results were therefore discarded. Thus, for stationary solutions with a spherical droplet inside the cubic cavity, the droplet shrinks while $\phi$ slightly increases both inside and outside the droplet. 

\section{Flow in a T-Shaped junction}  \label{ch_flow_T_junction}

\begin{figure}[htbp]
\centering
\includegraphics[width=\textwidth]{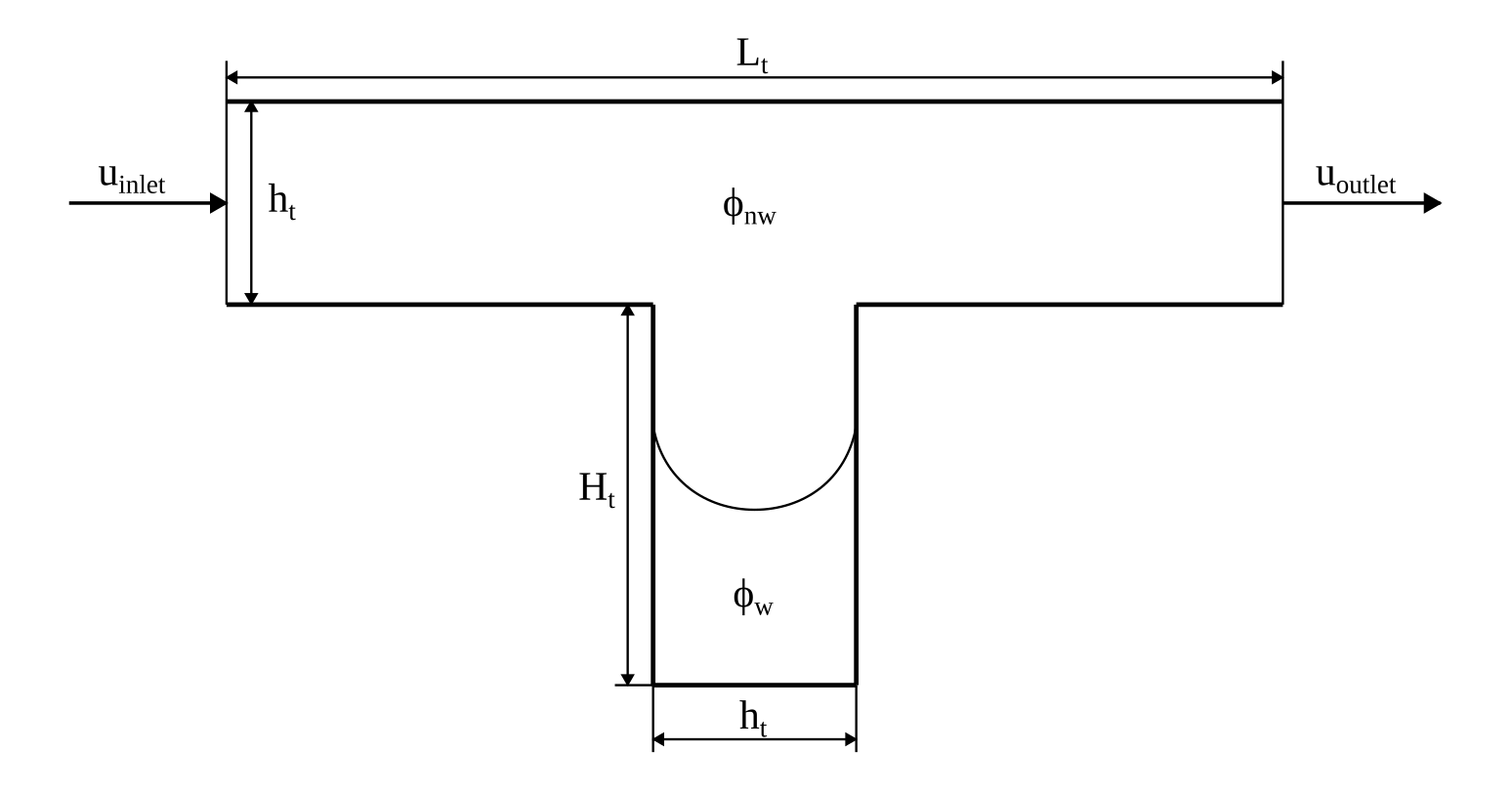}
\caption{Schematic of the T-Shaped junction geometry.}
\label{img:T_shape_scheme}
\end{figure}

Nonphysical artifacts associated with spurious phase generation are particularly evident in the benchmark problem involving a T-shaped junction, shown in \cref{img:T_shape_scheme}. The channels forming the junction have a square cross-section with side length $h_t = 15\,\mu m$, the length of the main flow path is $L_t = 60\,\mu m$, and the length of the dead end branch is $H_t = 30\,\mu m$. The computational domain was discretized using a uniform grid of cubic cells with edge length $h = 1\,\mu m$. A non-wetting phase is injected through the main channel at a prescribed velocity $u_{inlet} = 1\,mm/s$ at the inlet, with the same outflow condition at the outlet. The side branch is sealed with a no-flow boundary condition, trapping the wetting phase within this dead-end pore. The fluid properties and model parameters are identical to those used in the porous-medium simulations described previously. This configuration mimics a dead-end pore in a porous medium, which typically contains an immobile trapped phase. Accurate conservation of its volume is therefore essential for physically meaningful simulations.

\begin{figure}[htbp]
\centering
\includegraphics[width=\textwidth]{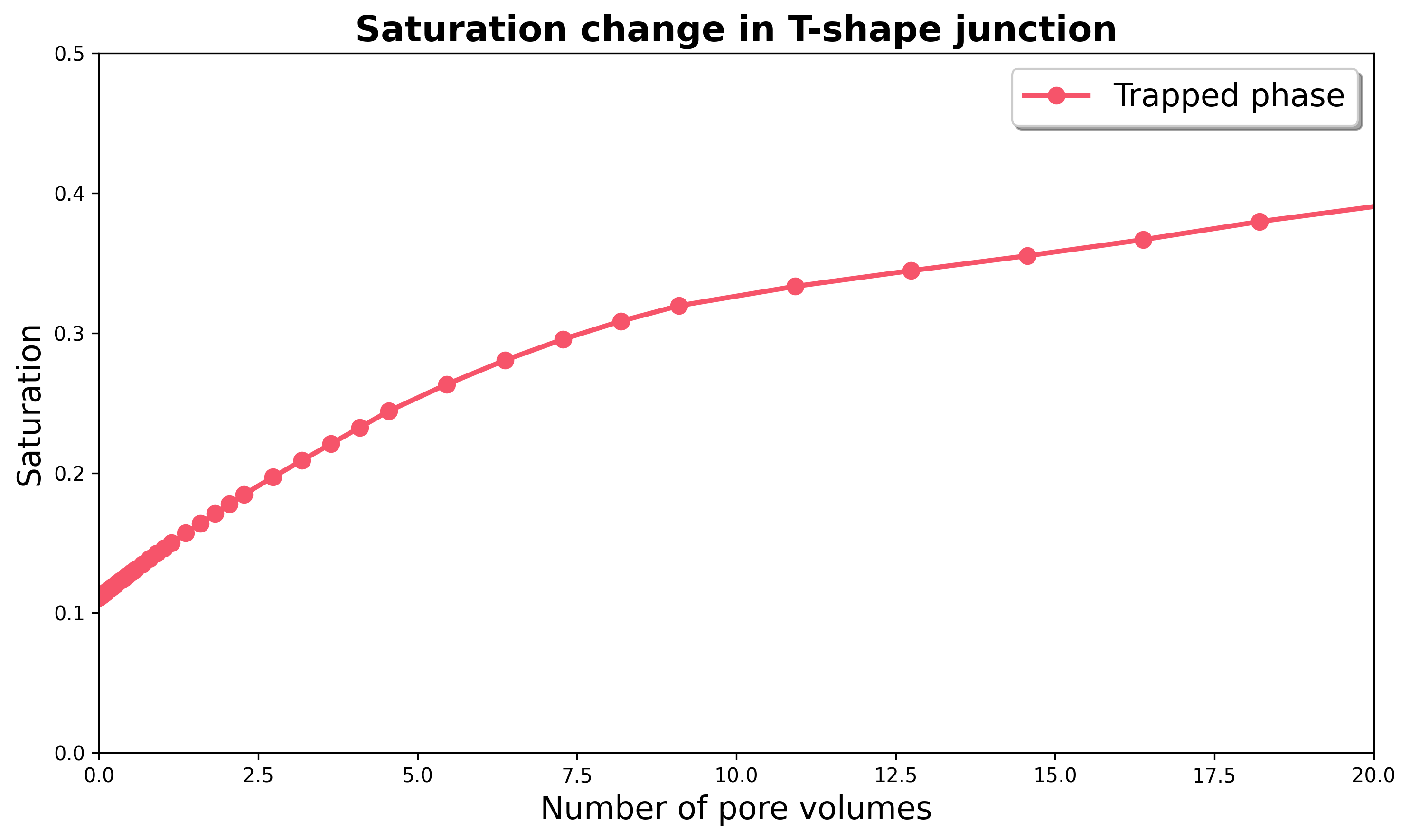}
\caption{Temporal evolution of the wetting-phase saturation in the T-shaped junction.}
\label{img:T_shape_saturation}
\end{figure}

\begin{figure}[htbp]
\centering
\includegraphics[width=\textwidth]{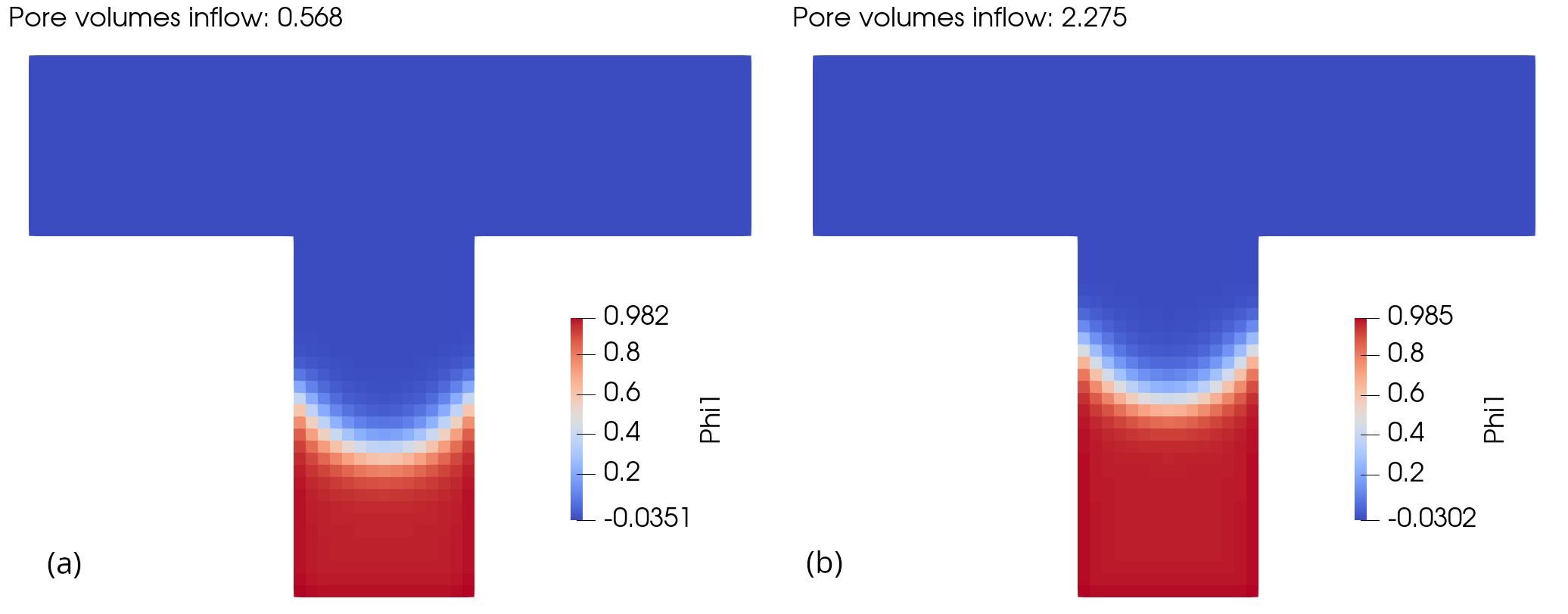}
\caption{Spatial distribution of the phase fraction in the T-shaped junction at two time instants.}
\label{img:T_shape_saturation_distr}
\end{figure}

\Cref{img:T_shape_saturation} shows the evolution of the wetting-phase saturation as a function of the number of pore volumes of non-wetting phase injected through the main channel, while \cref{img:T_shape_saturation_distr} displays the corresponding phase-fraction distributions at two representative time instants. The wetting-phase saturation increases by approximately 30\% relative to its initial value after the injection of just one pore volume of the non-wetting phase, clearly indicating spurious phase generation. In realistic displacement simulations through porous media, numerous dead-end pores are present throughout the computational domain, and approximately five global pore volumes of fluid must be injected before the system reaches a steady state. Given the significant disparity in scale, this injection protocol subjects each individual dead-end pore to hundreds of local pore volumes of flow passing through the adjacent main channel. Consequently, the artificial accumulation (or depletion) of phase mass in such regions can significantly distort the distribution of the trapped phase and compromise the fidelity of the overall simulation.

\section{Theoretical base of droplet shrinkage} \label{chap:drop_evaporation_theory}

\begin{figure}[htbp] 
\centering
\includegraphics[width=\textwidth]{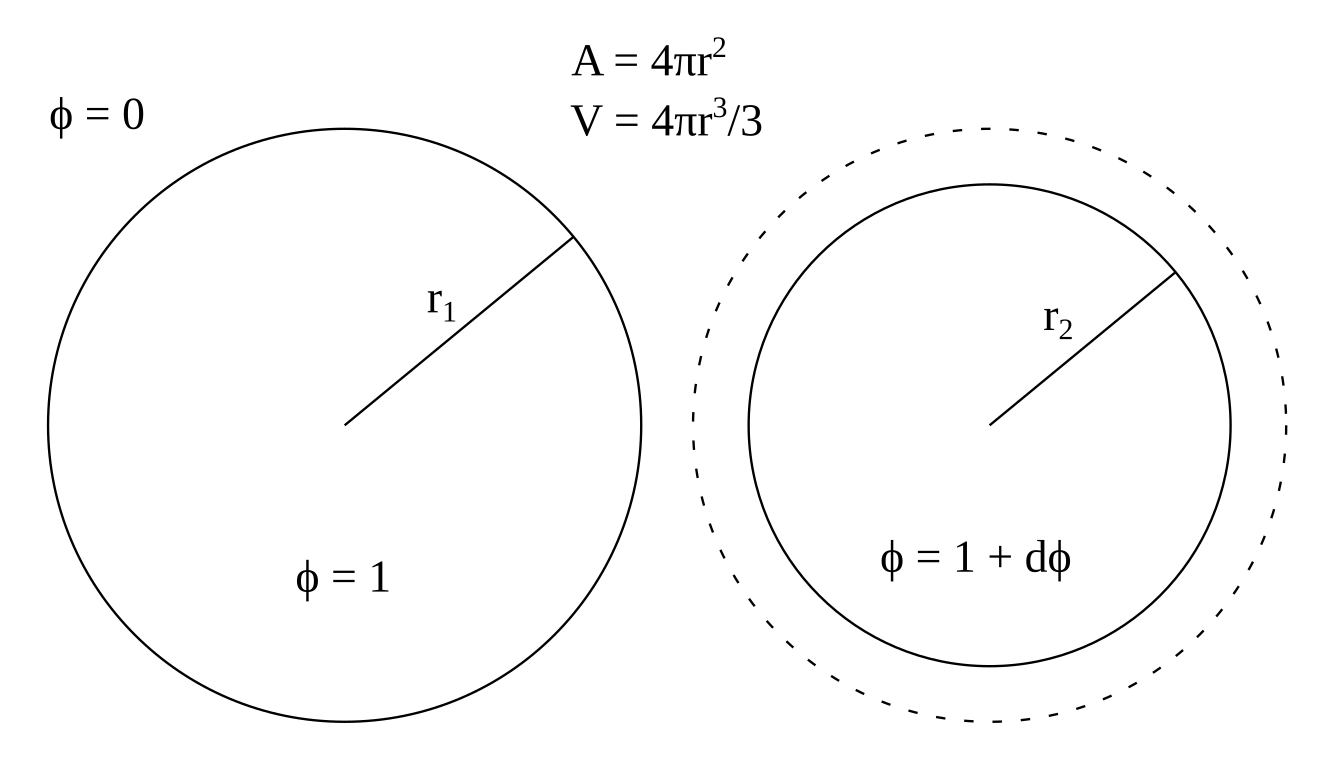}
\caption{Droplet shrinkage schematic}
\label{img:drop_shrinking_scheme}
\end{figure}

The theoretical basis for the numerical artifacts associated with droplet shrinkage in the diffuse interface method was established in \cite{YUE20071}. \Cref{img:drop_shrinking_scheme} schematically illustrates this shrinkage phenomenon.

We assume that the phase is uniformly distributed within the droplet and that the integral of the phase fraction over the entire computational domain remains constant in time. Initially, inside the droplet, the phase fraction is $\phi = 1$, and outside it is $\phi = 0$. The diffuse interface, where the phase fraction transitions from 1 to 0, has a characteristic width $\varepsilon$. We assume the initial droplet radius $r_1$ satisfies $r_1 \gg \varepsilon$. 

The energy of a droplet with radius $r_1$ and $\phi=1$ is not necessarily the global minimum in the diffuse interface formulation. The total free energy functional \eqref{energy functional two phase} consists of bulk and gradient contributions. The bulk energy density vanishes at $\phi=0$ and $\phi=1$. However, in the diffuse interface model, the system can lower the total energy by reducing the interfacial area (shrinkage of the droplet), even if this requires the internal phase fraction to exceed unity ($\phi > 1$). The equilibrium state is reached when the reduction in surface energy balances the increase in bulk energy. For sufficiently small droplets, it becomes energetically favorable to uniformly distribute the droplet phase over the entire computational domain, effectively eliminating the droplet while the total phase saturation remains constant. In the limit of an infinite computational domain, no stable droplet can exist. In the following derivation, we consider only the increase in the phase fraction inside the droplet and the decrease in the droplet radius. The critical radius for droplet disappearance is derived in \cite{YUE20071}.

Due to the minimization of interfacial energy, the droplet tends to shrink to a smaller radius $r_2 < r_1$. To satisfy global mass conservation, the phase fraction inside the droplet must increase by an amount $d\phi$. This deviation can be expressed as the ratio of the initial to final volumes
\begin{equation*}
    d\phi = \frac{r_1^3}{r_2^3} - 1.
\end{equation*}
Consequently, the phase fraction inside the shrunk droplet becomes $\phi = 1 + d\phi$.
Volume conservation requires
\begin{equation*}
(1 + d\phi) \frac{4}{3}\pi r_2^3 = \frac{4}{3}\pi r_1^3,
\end{equation*}
which yields the relationship between the radii
\begin{equation} \label{r2 dphi dependence}
r_2 = r_1 (1 + d\phi)^{-1/3}.
\end{equation}
We analyze the total free energy $F = F_\text{in} + F_s$, where $F_\text{in}$ is the bulk energy inside the droplet and $F_s$ is the interfacial energy. Using the energy functional \eqref{energy functional two phase} with the double-well potential $\Psi(\phi) = \phi^2(1-\phi)^2$, the bulk energy density at $\phi = 1 + d\phi$ is:
\begin{equation*}
12 \frac{\sigma}{\varepsilon} \Psi(1 + d\phi) 
= 12 \frac{\sigma}{\varepsilon} (1 + d\phi)^2 (d\phi)^2.
\end{equation*}
Integrating this density over the droplet volume $V_2 = \frac{4}{3}\pi r_2^3$ and substituting \eqref{r2 dphi dependence} gives
\begin{align*}
F_{\text{in}} 
&= 12 \frac{\sigma}{\varepsilon} (1 + d\phi)^2 (d\phi)^2 \frac{4}{3}\pi r_2^3 \\
&= 12 \frac{\sigma}{\varepsilon} (1 + d\phi)^2 (d\phi)^2 \frac{4}{3}\pi r_1^3 (1 + d\phi)^{-1} \\
&= 16 \pi \frac{\sigma}{\varepsilon} r_1^3 (1 + d\phi) (d\phi)^2  \\
&= 16 \pi \frac{\sigma}{\varepsilon} r_1^3 \left( d\phi^2 + d\phi^3 \right).
\end{align*}
Retaining only the leading-order term (since $d\phi^3 \ll d\phi^2$ for small deviations) yields
\begin{equation} \label{Fin final}
F_{\text{in}} \approx 16\pi \, \frac{\sigma}{\varepsilon} \, r_1^3 \, d\phi^2.
\end{equation}
The interfacial energy for a spherical droplet is given by $F_s = \sigma \cdot 4\pi r_2^2$, where $\sigma$ is the surface tension \cite{Cahn_Hilliard_two_phase_0_1}. Substituting \eqref{r2 dphi dependence}
\begin{equation} \label{Fs final}
F_s = 4\pi \sigma r_1^2 (1 + d\phi)^{-2/3}.
\end{equation}
The total energy $F(d\phi) = F_{\text{in}} + F_s$ is minimized at equilibrium. Differentiating \eqref{Fin final} and \eqref{Fs final} with respect to $d\phi$ and setting the derivative to zero yields
\begin{equation*}
32\pi \, \frac{\sigma}{\varepsilon} \, r_1^3 \, d\phi 
\;-\; \frac{8\pi}{3} \sigma r_1^2 (1 + d\phi)^{-5/3} = 0.
\end{equation*}
Since we anticipate $d\phi = O(\varepsilon/r_1) \ll 1$, we approximate $(1 + d\phi)^{-5/3} \approx 1$. Solving for $d\phi$
\begin{equation} \label{dphi result}
d\phi = \frac{1}{12} \, \frac{\varepsilon}{r_1}.
\end{equation}
As the difference between the initial radius $r_1$ and the equilibrium radius $r_2$ is of order $O(\varepsilon/r_1)$, they differ only slightly. Therefore, in practical calculations, the observed equilibrium radius $r_2$ is used in place of $r_1$ on the right-hand side of \cref{dphi result}.

\section{Correction of the double-well potential}

\begin{figure}[htbp]
\centering
\includegraphics[width=\textwidth]{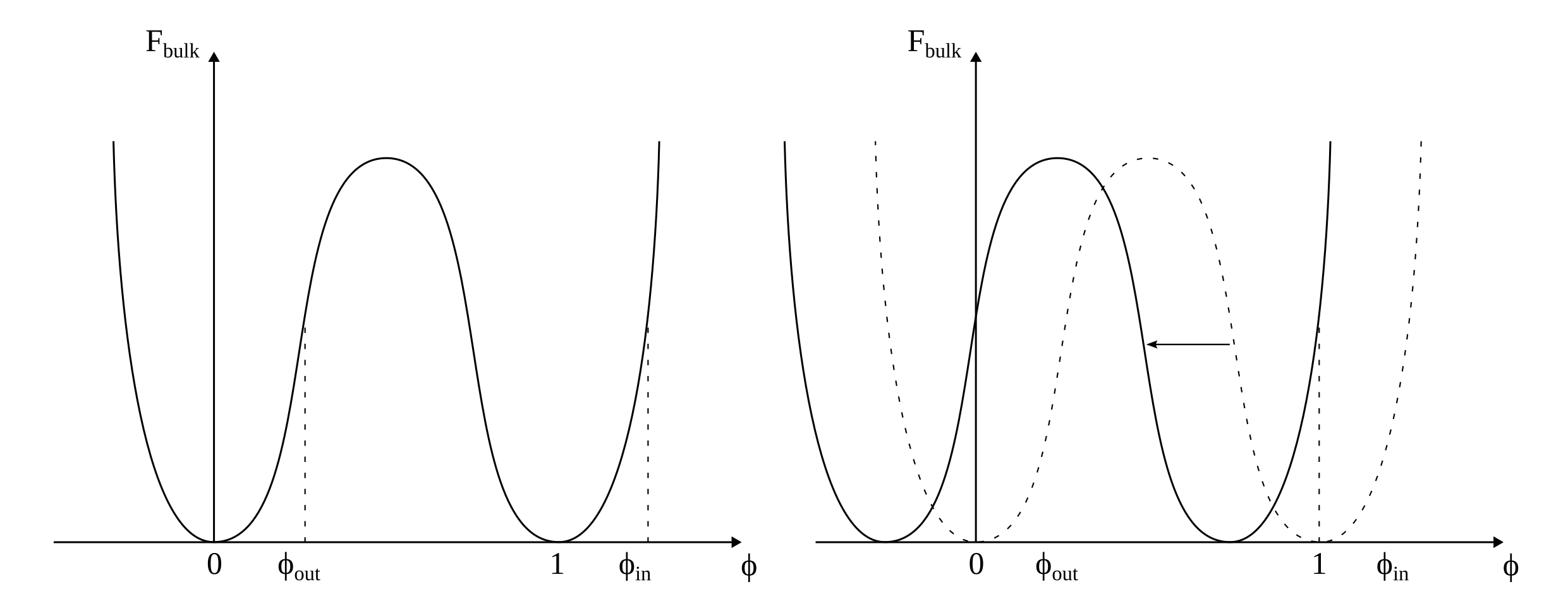}
\caption{Schematic illustration of the double-well potential shift.}
\label{Potential_shift_pic}
\end{figure}

\Cref{box_without_fix} reveals that the phase fraction exhibits an almost uniform shift in one direction, resulting in values that exceed the desired equilibrium bulk values both inside and outside the droplet. A straightforward mitigation strategy is to shift the double-well potential in the opposite direction, thereby restoring the order parameter toward its equilibrium values of 0 and 1. This approach is illustrated in \cref{Potential_shift_pic}.

For a single droplet, a uniform shift is sufficient, but a more general approach is needed for systems containing multiple droplets of different sizes. This shift must be applied pointwise across the domain, requiring a spatially varying correction field $\phi_s$. Using \cref{dphi result}, $d\phi$ can be estimated locally if the local droplet radius $r_2$ is known. The radius is obtained from the curvature $\kappa$ via $r_2 = 2 / \kappa$, where the curvature is computed from the phase field according to

\begin{equation} \label{curvature_eqn}
\kappa = \nabla \cdot \left ( \frac{\nabla \phi}{\left | \nabla \phi \right |} \right ).
\end{equation}
In expanded form, this expression becomes

\begin{subequations}
\begin{align}
\kappa = &
\frac{\left(\phi_x^2+\phi_y^2+\phi_z^2\right)
    \left(\phi_{xx}+\phi_{yy}+\phi_{zz}\right) }{\left(\phi_x^2+\phi_y^2+\phi_z^2\right)^{\frac{3}{2}}} \\
- &\frac{ \left(\phi_x\right)^2 \phi_{xx}
    + \phi_y^2 \phi_{yy}
    + \phi_z^2 \phi_{zz}
    + 2 \phi_x \phi_y \phi_{xy}
    + 2 \phi_x \phi_z \phi_{xz}
    + 2 \phi_y \phi_z \phi_{yz} }{\left(\phi_x^2+\phi_y^2+\phi_z^2\right)^{\frac{3}{2}}}.
\end{align}
\end{subequations}
This expression yields the curvature in the interfacial region, which is then used to compute the local shift magnitude. 

To compute a field $\phi_s$ that varies smoothly over the entire domain and remains close to $d\phi$ from \cref{dphi result} at the interfaces, we solve a penaltized Laplace equation:
\begin{equation} \label{shift_laplace}
\Delta \phi_s + \left ( \nabla \phi \right )^2 \left( \phi_s - \frac{\varepsilon \kappa}{24} \right) = 0,
\end{equation}
with zero normal gradient boundary condition
\begin{equation}
\vec{n} \cdot \nabla \phi_s = 0.
\end{equation}
The term $|\nabla \phi|^2$ is large at the interface and vanishes in the bulk phases. Therefore, it enforces $\phi_s \approx \varepsilon \kappa / 24 = d\phi$ at the interface, while $\phi_s$ satisfies Laplace equation away from the interface. The resulting field $\phi_s$ is then used to shift the bulk energy density:
\begin{equation} \label{shrink_fix_eq}
\mu = 12 \frac{\sigma}{\varepsilon} \Psi'(\phi - \phi_s) - \frac{3}{2} \sigma \varepsilon \Delta \phi.
\end{equation}
The shift $\phi_s$ is not included in the implicit system matrix. Instead, it is computed explicitly from the phase field $\phi^n$ at the previous time step. 

Thus, the proposed correction is applied as follows. We use \eqref{shrink_fix_eq} in place of \eqref{chem_potential_eq}. Its discretization is straightforward because $\phi_s$ is cell-centered and is computed using variables from the previous time step:
\begin{equation*}
\mu_{i,j,k} = 12 \frac{\sigma}{\varepsilon} \Psi'(\phi_{i,j,k} - (\phi_s)_{i,j,k}) - \frac{3}{2} \sigma \varepsilon \left (\Delta_h \phi\right )_{i,j,k}.
\end{equation*}

\section{Numerical validation of the correction scheme}

The previously described correction was implemented, and test simulations were conducted for three previously described scenarios: flow through a porous medium, a droplet in a cubic cavity, and a T-shaped junction.

\begin{table}[htbp]
\begin{tabular}{|c|c|c|c|c|}
\hline
Case & \(L\), $\mu m$ & \(h_d\), $\mu m$ & \(d\), $\mu m$ & \(\phi_{max}\), \(\phi_{min}\) \\
\hline
1 & 30 & 18 & 21 & 0.99626, 0.00146671 \\
2 & 60 & 20 & 23 & 0.99695, 0.00239898 \\
3 & 60 & 25 & 27 & 0.99794, 0.00162382 \\
4 & 60 & 30 & 35 & 0.99862, 0.00108721 \\
5 & 60 & 35 & 41 & 0.99889, 0.00086623 \\
6 & 60 & 40 & 49 & 0.99913, 0.00065861 \\
7 & 60 & 45 & -- & -- \\
8 & 80 & 30 & 35 & 0.99861, 0.00108875 \\
9 & 80 & 35 & 41 & 0.99888, 0.00086809 \\
10 & 80 & 40 & 49 & 0.99914, 0.00065462 \\
11 & 80 & 45 & 53 & 0.99925, 0.00056227 \\
12 & 80 & 50 & 61 & 0.99938, 0.00046119 \\
13 & 80 & 55 & 65 & 0.99945, 0.00038403 \\
14 & 80 & 60 & -- & -- \\
\hline
\end{tabular}
\caption{Simulation parameters and results for a droplet in a cubic cavity with the proposed correction.}
\label{box_with_fix}
\end{table}

\begin{figure}[htbp]
\centering
\includegraphics[width=\textwidth]{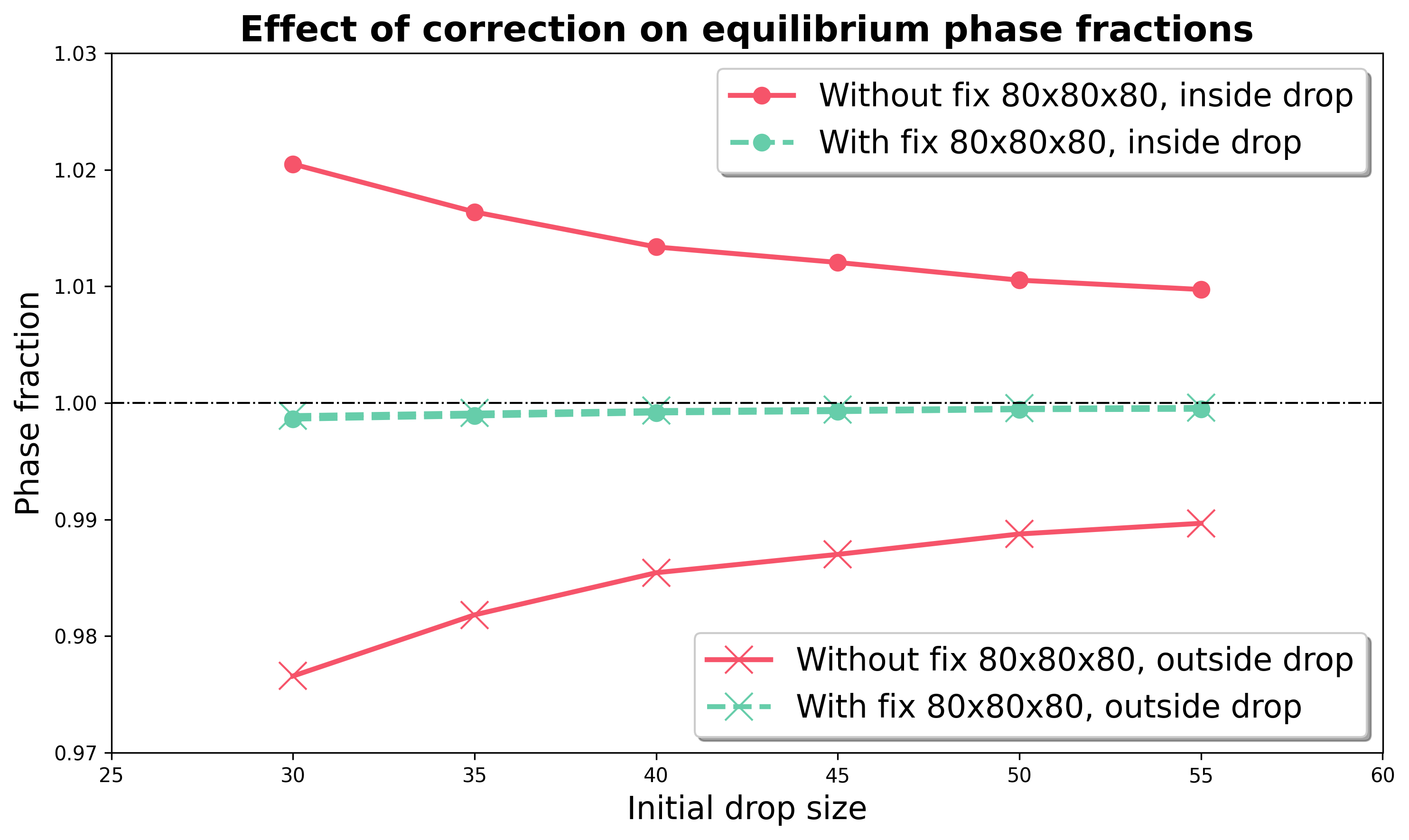}
\caption{Effect of the correction on equilibrium phase fractions.}
\label{shrink_correction_plot}
\end{figure}

\Cref{box_with_fix} presents the results obtained with the correction defined in \cref{shrink_fix_eq}. The phase fraction now remains strictly within the physical bounds 0 and 1. Notably, in Case~2, where the droplet vanished in the uncorrected simulation, it now persists. For clarity, data from \cref{box_without_fix,box_with_fix} are summarized in \cref{shrink_correction_plot}, which shows the equilibrium value of phase~1 inside the droplet (circles) and phase~2 outside (crosses). With the correction, the phase fractions inside and outside the droplet are significantly closer to their physical values.

In the T-shaped junction test case described in \cref{ch_flow_T_junction}, the proposed correction greatly suppresses nonphysical effects.
\begin{figure}[htbp]
\centering
\includegraphics[width=\textwidth]{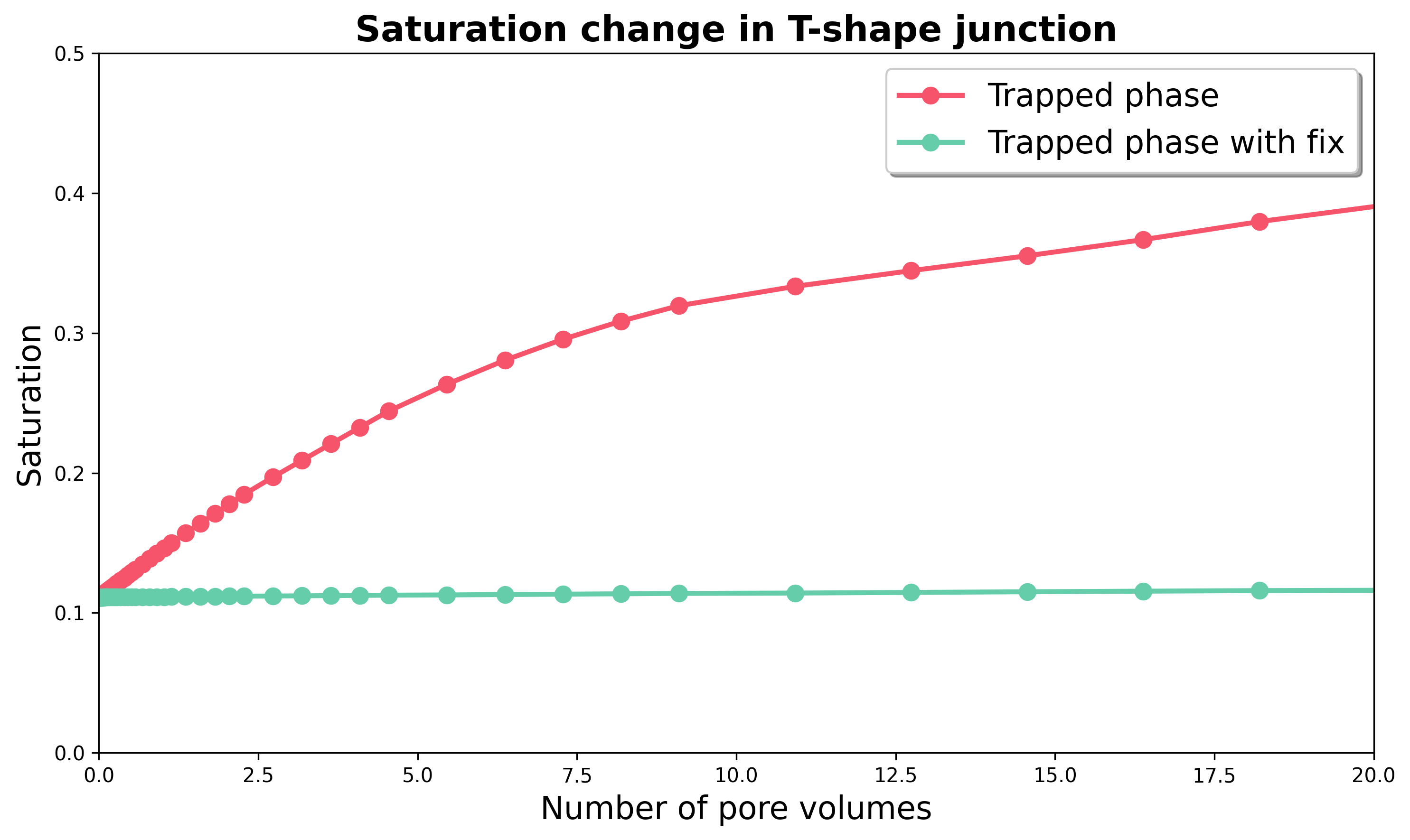}
\caption{Saturation of the wetting phase in a T-shaped junction.}
\label{img:T_shape_saturation_fix}
\end{figure}
\Cref{img:T_shape_saturation_fix} shows the evolution of the wetting-phase saturation as a function of the number of pore volumes of non-wetting phase injected through the junction, comparing results with and without the correction \cref{shrink_fix_eq}. With the proposed correction, the saturation change is limited to 0.3\% per injected pore volume, which is approximately two orders of magnitude lower than the uncorrected case. However, a gradual drift remains and at 20 pore volumes, the relative change reaches approximately 4.6\%, which may still influence the residual saturation over long time scales.

The porous medium flow case also demonstrates significant improvement.
\begin{figure}[htbp]
\centering
\includegraphics[width=\textwidth]{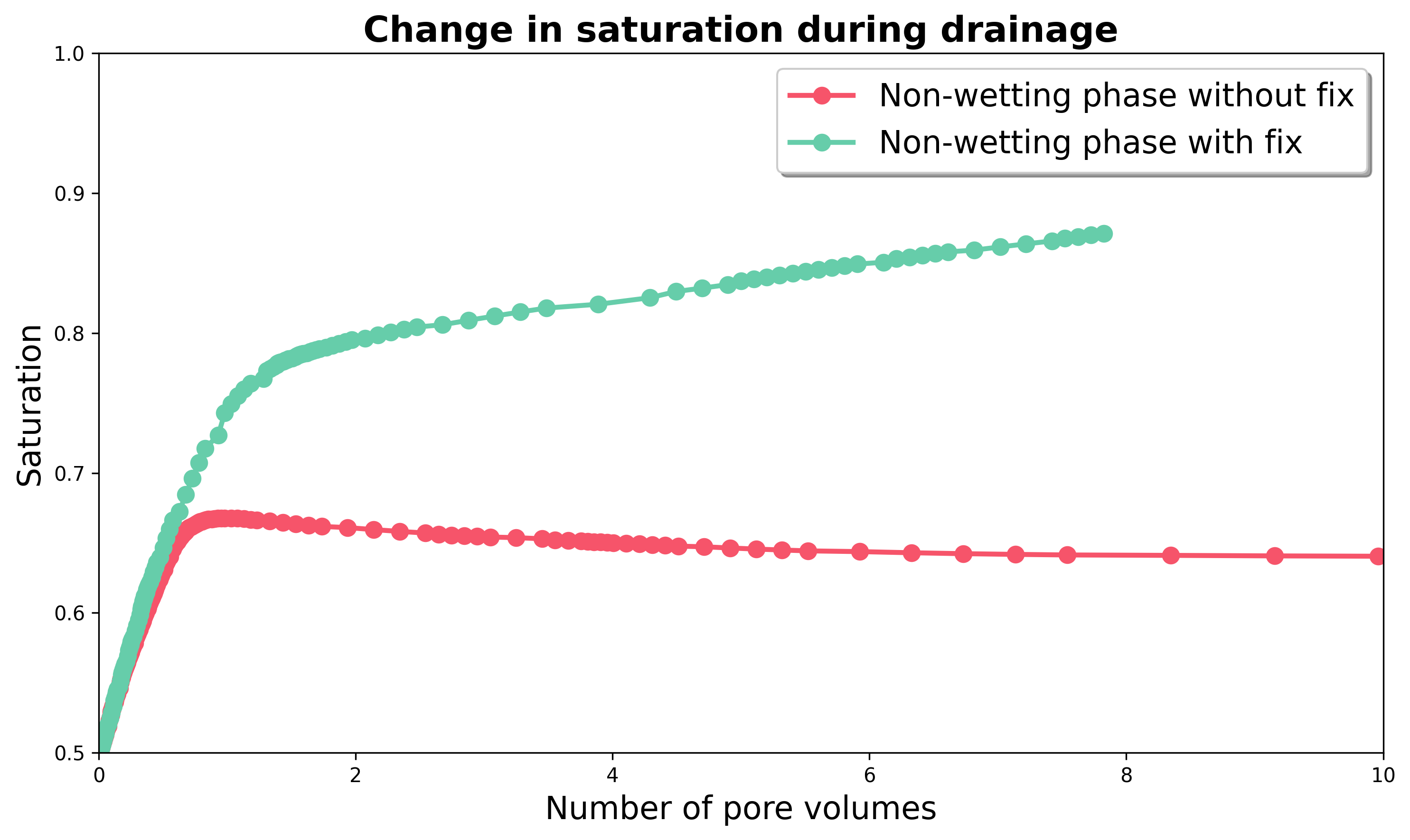}
\caption{Saturation evolution during drainage with the correction \cref{shrink_fix_eq}}
\label{img:Saturation_plot_fix}
\end{figure}
\Cref{img:Saturation_plot_fix} displays the effect of the correction on the non-wetting phase saturation. Although the potential shift is only about 5\%, it profoundly affects the displacement dynamics. The curves diverge before one pore volume is injected. The uncorrected case exhibits nonphysical behavior: after the first pore volume, the non-wetting phase saturation decreases while the wetting phase increases, despite only non-wetting fluid being injected. In contrast, the corrected case shows a gradual, physical removal of the wetting phase. 
\begin{figure}[htbp]
\centering
\includegraphics[width=\textwidth]{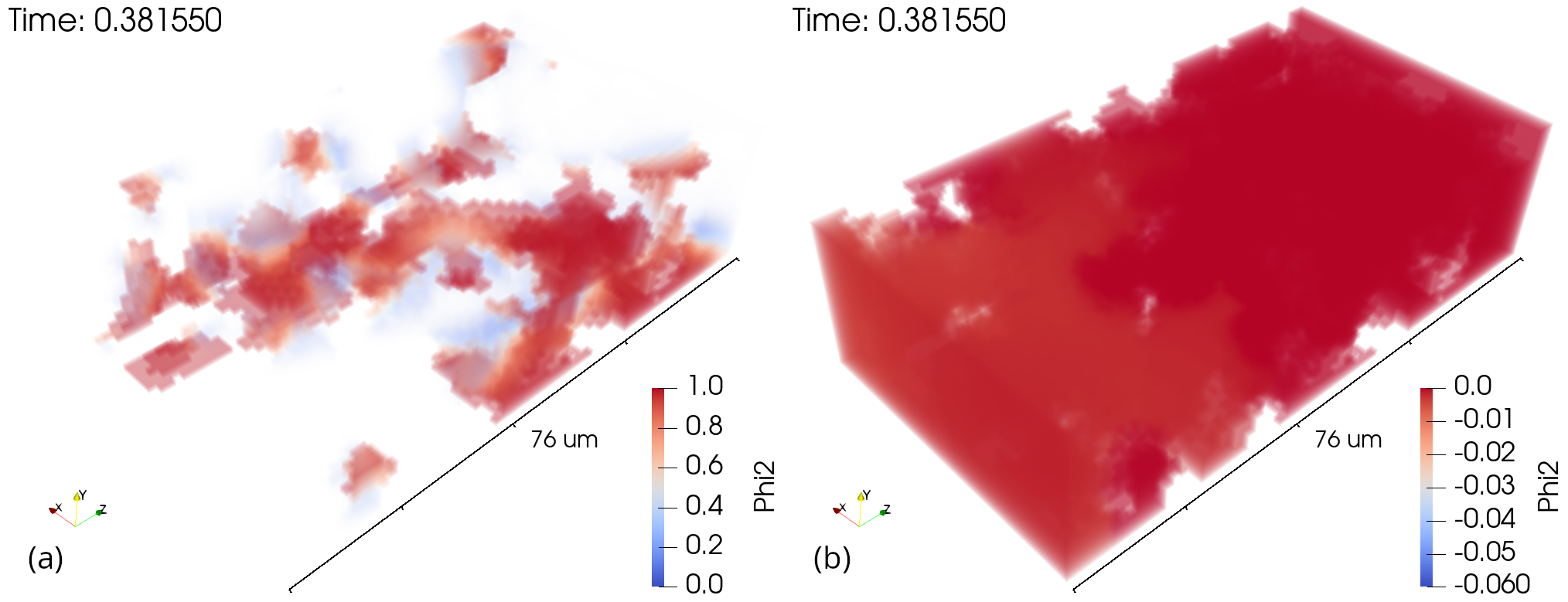}
\caption{Wetting phase distribution in the porous medium with the correction \cref{shrink_fix_eq}: (a) $\phi$ distribution within physical bounds, (b) $\phi$ distribution in the range -0.06 to 0.}
\label{img:porous_med_distribution_fix}
\end{figure}
\Cref{img:porous_med_distribution_fix} illustrates the effect of the correction on the wetting phase distribution. In \Cref{img:porous_med_distribution_fix} (a), it is evident that there is significantly less wetting phase in the computational domain compared to \Cref{img:porous_med_distribution}(a) as the artificial accumulation of it was suppressed. Similarly, \cref{img:porous_med_distribution_fix} (b) shows the wetting phase saturation near the outlet is much closer to zero than in the uncorrected case shown in \Cref{img:porous_med_distribution} (b).

\section{Conclusions}

Building on the work of \cite{YUE20071}, a correction for the two-phase Cahn--Hilliard equations was developed and implemented. This correction mitigates nonphysical effects that arise in the modeling of immiscible fluids. It requires solving an additional Laplace-type equation and computing the interface curvature using \cref{curvature_eqn}. The effectiveness of the correction was evaluated on three test cases: a droplet in a cubic computational domain, wetting-phase displacement in a porous medium, and flow through a T-shaped junction. The results demonstrate that the correction significantly reduces nonphysical artifacts, although it does not eliminate them entirely. Nevertheless, the correction is straightforward to implement and does not appreciably increase the complexity of the numerical solution.

\printbibliography

\end{document}